\documentclass[epj-spec]{svjour}

\usepackage{amsmath,amssymb,bbm}

\newcommand{\mathd}{\mathrm{d}}
\newcommand{\mathe}{\mathrm{e}}

\newcommand{\tmmathbf}[1]{\ensuremath{\boldsymbol{#1}}}
\newcommand{\tmop}[1]{\ensuremath{\operatorname{#1}}}
\newcommand{\tmtextbf}[1]{{\bfseries{#1}}}
\newcommand{\tmtextit}[1]{{\itshape{#1}}}

\begin{document}

\author{Bassano Vacchini\inst{1}\fnmsep\thanks{\email{bassano.vacchini@mi.infn.it}}
\and Francesco
  Petruccione\inst{2}\fnmsep\thanks{\email{petruccione@ukzn.ac.za}}}

\institute{{Dipartimento di Fisica
 dell'Universit\`a di Milano and INFN,
 Sezione di Milano
 Via Celoria 16, 20133, Milan, Italy} \and {School of Physics, Quantum Research Group, 
   University of KwaZulu-Natal, Westville Campus, Durban 4000, South
   Africa}}



\title{Kinetic description of quantum Brownian motion}

\abstract{
  We stress the relevance of the two features of translational invariance and
  atomic nature of the gas in the quantum description of the motion of a
  massive test particle in a gas, corresponding to the original picture of
  Einstein used in the characterization of Brownian motion. The master
  equation describing the reduced dynamics of the test particle is of Lindblad
  form and complies with the requirement of covariance under translations.
}\maketitle

\section{Introduction}\label{sec:introduction}

In recent times there has been growing interest in the study of open quantum
systems {\cite{Breuer2007}}, driven by researches on both applications and
foundations of quantum theory. Indeed the community of researchers involved in
the subject, in spite of the precise label attached to it, ranges from
physicists and chemists, to mathematicians and probabilists. As it often
happens the richness of the subject allows for different approaches and biases
in the treatment of the same problem. In the present paper we come back to the
quest for a deep understanding of one of the apparently simplest, but also
paradigmatic situations in open quantum system theory, that is to say the
motion of a quantum massive test particle in a homogeneous gas. Two guiding
ideas, also present in the classical approach to Brownian motion of a
suspended particle by Einstein, appear in the treatment: symmetry under
translations and discrete nature of matter. Following work done in
{\cite{Petruccione2005a}} we will see that these two concepts reflect
themselves in the mathematical properties of the mapping describing the
reduced dynamics, which has to be covariant under translations, as well as in
the physical meaning of the two-point correlation function of the gas
appearing in the expression of the master equation, which expresses the
density fluctuations in the gas.

\section{Translational Invariance\label{sec:ti}}

\subsection{Microscopic Hamiltonian}\label{sec:transl-invar}

As a first step we characterize the general structure of microscopic
Hamiltonians accounting for a translationally invariant reduced dynamics for
the test particle. We consider a test particle subject to a translationally
invariant interaction with a homogeneous bath, with a potential at most
linearly depending on position, e.g., a constant gravitational field. The
microscopic Hamiltonian may be written as usual
\begin{equation}
  \label{eq:1} H = H_S + H_B + V,
\end{equation}
where the subscripts $S$ and $B$ stand for system and bath respectively, while
$H_S$ and $H_B$ satisfy the aforementioned constraints. The interaction relies
on a translationally invariant potential and is of the general form
\begin{equation}
  \label{eq:2} V = \int d^3 \hspace{-0.25em} \tmmathbf{x} \int d^3
  \hspace{-0.25em} \tmmathbf{y} \hspace{0.25em} A_S (\tmmathbf{x}) V
  (\tmmathbf{x}-\tmmathbf{y}) A_B (\tmmathbf{y}),
\end{equation}
where $A_S (\tmmathbf{x})$ and $A_B (\tmmathbf{y})$ are self-adjoint operators
of system and bath respectively, expressing the coupling between the two. The
invariance under translations of the potential allows to express Eq.
(\ref{eq:2}) in a very simple way in terms of the Fourier transformed
quantities according to
\begin{equation}
  \label{eq:6} V = \int d^3 \hspace{-0.25em} \tmmathbf{Q} \hspace{0.25em}
  \tilde{V} (\tmmathbf{Q}) A_S (\tmmathbf{Q}) A_B^{\dagger} (\tmmathbf{Q}) .
\end{equation}
In order to focus on a quantum description of Brownian motion we consider a
density-density coupling, so that $A_{S / B} (\tmmathbf{x}) = N_{S / B}
(\tmmathbf{x})$, with $N_{S / B} (\tmmathbf{x})$ the number-density operator
of system and bath respectively, whose Fourier transform $\rho_{\tmmathbf{Q}}$
\begin{eqnarray}
  \rho_{\tmmathbf{Q}} & \equiv & \int d^3 \hspace{-0.25em} \tmmathbf{x}
  \hspace{0.25em} e^{- \frac{i}{\hbar} \tmmathbf{Q} \cdot \tmmathbf{x}} N_B
  (\tmmathbf{x})  \label{eq:15}
\end{eqnarray}
is also called $\tmmathbf{Q}$-component of the number-density operator
{\cite{Lovesey1984,Pitaevskii2003}}. Eq. (\ref{eq:6}) thus becomes
\begin{equation}
  \label{eq:16} V = \int d^3 \hspace{-0.25em} \tmmathbf{Q} \hspace{0.25em}
  \tilde{V} (\tmmathbf{Q}) A_S (\tmmathbf{Q}) \rho_{\tmmathbf{Q}}^{\dagger}
  (\tmmathbf{Q}) .
\end{equation}
Note that an interaction of the form (\ref{eq:16}), besides being
translationally invariant, commutes with the number operators $N_S$ and $N_B$,
so that the elementary interaction events do bring in exchanges of momentum
between the test particle and the environment, but the number of particles or
quanta in both systems are independently conserved, thus typically describing
an interaction in terms of collisions.

\subsection{Quantum linear Boltzmann equation\label{sec:qlbe}}

The case of density-density coupling given by (\ref{eq:16}), when the
reservoir is a free quantum gas, has been dealt with in
{\cite{Vacchini2000a,Vacchini2001a,Vacchini2001b}}, and the relevant test
particle correlation function turns out to be the so-called dynamic structure
factor {\cite{Lovesey1984,Pitaevskii2003}}
\begin{equation}
  \label{eq:37} S (\tmmathbf{Q}, E) = \frac{1}{2 \pi \hbar} \frac{1}{N} \int
  dt \hspace{0.25em} e^{\frac{i}{\hbar} Et} \langle
  \rho_{\tmmathbf{Q}}^{\dagger} \rho_{\tmmathbf{Q}} (t) \rangle,
\end{equation}
where contrary to the usual conventions, momentum and energy are considered to
be positive when transferred to the test particle, on which we are now
focusing our attention, rather than on the macroscopic system. The master
equation then takes the form
\begin{eqnarray}
  \frac{\mathd \varrho}{dt} & = & - \frac{i}{\hbar} [ \mathsf{H}_0, \varrho] 
  \label{eq:39}\\
  &  & + \frac{2 \pi}{\hbar} (2 \pi \hbar)^3 n_{\tmop{gas}} \int d^3
  \hspace{-0.25em} \tmmathbf{Q} \hspace{0.25em} | \tilde{V} (\tmmathbf{Q}) |^2
  \left[ e^{\frac{i}{\hbar} \tmmathbf{Q} \cdot \mathsf{X}} \sqrt{S
  (\tmmathbf{Q}, E (\tmmathbf{Q}, \mathsf{P}))} \varrho \sqrt{S (\tmmathbf{Q},
  E (\tmmathbf{Q}, \mathsf{P}))} e^{- \frac{i}{\hbar} \tmmathbf{Q} \cdot
  \mathsf{X}} \right. \nonumber\\
  &  & \left. \phantom{\left. - \frac{1}{2} \left\{ L^{\dag} \left(
  \tmmathbf{p}, \mathsf{P} ; \tmmathbf{Q} \right) L \left( \tmmathbf{p},
  \mathsf{P} ; \tmmathbf{Q} \right), \varrho \right\} \right], S
  (\tmmathbf{Q}, E (\tmmathbf{Q}, \mathsf{P}))} - \frac{1}{2} \left\{ S
  (\tmmathbf{Q}, E (\tmmathbf{Q}, \mathsf{P})), \varrho \right\} \right],
  \nonumber
\end{eqnarray}
where $\mathsf{H}_0$ is the free particle Hamiltonian, $\mathsf{X}$ and
$\mathsf{P}$ position and momentum operator for the test particle,
$n_{\tmop{gas}}$ the density of the homogeneous gas, and the dynamic structure
factor appears operator-valued: in fact the energy transfer in each collision,
which is given by
\begin{equation}
  \label{eq:40} E (\tmmathbf{Q}, \tmmathbf{P}) = \frac{( \mathbf{\tmmathbf{P}}
  +\tmmathbf{Q})^2}{2 M} - \frac{\mathbf{\tmmathbf{P}}^2}{2 M},
\end{equation}
with $M$ the mass of the test particle, is turned into an operator by
replacing $\tmmathbf{P}$ with $\mathsf{P}$. For the case of a free gas of
particles obeying Maxwell-Boltzmann statistics the dynamic structure factor
takes the explicit form
\begin{equation}
  \label{eq:41} S_{} (\tmmathbf{Q}, E) = \sqrt{\frac{\beta m}{2 \pi}}
  \frac{1}{Q} e^{- \frac{\beta}{8 m} \frac{\left( 2 mE + Q^2
  \right)^2}{Q^2}^{}}
\end{equation}
with $\beta$ the inverse temperature and $m$ the mass of the gas particles.

This result has been confirmed in recent, more general work
{\cite{Hornberger2006b,Hornberger2007c}}, not relying on the Born
approximation, where instead of the Fourier transform of the interaction
potential the full scattering amplitude describing the collisions between test
particle and gas particles appear operator-valued. In the general case the
master equation describing the collisional dynamics takes the following form
\begin{eqnarray}
  \frac{\mathd \varrho}{\tmop{dt}} & = & - \frac{i}{\hbar} \left[
  \mathsf{H}_0, \varrho \right]  \label{eq:full}\\
  &  & + \int d^3 \hspace{-0.25em} \tmmathbf{Q} \int_{\tmmathbf{Q}^{\bot}}
  d^3 \hspace{-0.25em} \tmmathbf{p} \left[ \mathe^{i\tmmathbf{Q} \cdot
  \mathsf{X} / \hbar} L \left( \tmmathbf{p}, \mathsf{P} ; \tmmathbf{Q} \right)
  \varrho L^{\dag} \left( \tmmathbf{p}, \mathsf{P} ; \tmmathbf{Q} \right)
  \mathe^{- i\tmmathbf{Q} \cdot \mathsf{X} / \hbar} \right. \nonumber\\
  &  & \left. \phantom{\left. - \frac{1}{2} \left\{ L^{\dag} \left(
  \tmmathbf{p}, \mathsf{P} ; \tmmathbf{Q} \right) L \left( \tmmathbf{p},
  \mathsf{P} ; \tmmathbf{Q} \right), \varrho \right\} \right],} - \frac{1}{2}
  \left\{ L^{\dag} \left( \tmmathbf{p}, \mathsf{P} ; \tmmathbf{Q} \right) L
  \left( \tmmathbf{p}, \mathsf{P} ; \tmmathbf{Q} \right), \varrho \right\}
  \right], \nonumber
\end{eqnarray}
where the Lindblad operators $L \left( \tmmathbf{p}, \tmmathbf{P} ;
\tmmathbf{Q} \right)$ are given by
\begin{eqnarray}
  L \left( \tmmathbf{p}, \tmmathbf{P} ; \tmmathbf{Q} \right) & = &
  \sqrt{\frac{n_{\tmop{gas}} m}{m_{\ast}^2 Q}} f \left( \tmop{rel} \left(
  \tmmathbf{p}_{\bot \tmmathbf{Q}}, \tmmathbf{P}_{\bot \tmmathbf{Q}} \right) -
  \frac{\tmmathbf{Q}}{2}, \tmop{rel} \left( \tmmathbf{p}_{\bot \tmmathbf{Q}},
  \tmmathbf{P}_{\bot \tmmathbf{Q}} \right) + \frac{\tmmathbf{Q}}{2} \right) 
  \label{eq:L}\\
  &  & \times \sqrt{\mu_{_{\tmop{MB}}} \left( \tmmathbf{p}_{\bot
  \tmmathbf{Q}} + \frac{m}{m_{\ast}} \frac{\tmmathbf{Q}}{2} + \frac{m}{M}
  \tmmathbf{P}_{\| \tmmathbf{Q}} \right)}^{}, \nonumber
\end{eqnarray}
involving the elastic scattering amplitude $f \left( \tmmathbf{p}_f,
\tmmathbf{p}_i) \right.$ and the momentum distribution function
$\mu_{_{\tmop{MB}}} \left( \tmmathbf{p} \right)$ of the gas momenta given by
the Maxwell-Boltzmann expression. We have further denoted relative momenta as
$\tmop{rel} \left( \tmmathbf{p}, \tmmathbf{P} \right) \equiv \left( m_{\ast} /
m \right) \tmmathbf{p} - \left( m_{\ast} / M \right) \tmmathbf{P}$, with
$m_{\ast}$ the reduced mass, while the subscripts $\| \tmmathbf{Q}$ and $\perp
\tmmathbf{Q}$ indicate the component of a vector (or operator)
$\tmop{parallel}$ and perpendicular to the momentum transfer $\tmmathbf{Q}$,
so that $\tmmathbf{P}_{\| \tmmathbf{Q}} = \left( \tmmathbf{P} \cdot
\tmmathbf{Q} \right) \tmmathbf{Q} / Q^2$ and $\tmmathbf{P}_{\bot \tmmathbf{Q}}
= \tmmathbf{P} - \tmmathbf{P}_{\| \tmmathbf{Q}}$ respectively. Exploiting the
identity
\begin{eqnarray}
  \frac{m}{Q} \mu_{_{\tmop{MB}}} \left( \tmmathbf{p} \mathbf{}_{\perp
  \tmmathbf{Q}} \text{$+ \frac{m}{m_{\ast}^{}} \frac{\tmmathbf{Q}}{2} +
  \frac{m}{M} \tmmathbf{P}_{\| \tmmathbf{Q}}$} \right) & = & \frac{m}{Q}
  \mu_{_{\tmop{MB}}} \left( \tmmathbf{p} \mathbf{}_{\perp \tmmathbf{Q}}
  \text{$+ \left( \frac{2 mE \left( \tmmathbf{Q}, \tmmathbf{P} \right) +
  Q^2}{Q^2} \right) \frac{\tmmathbf{Q}}{2}$} \right)  \label{eq:equiv}\\
  & = & \mu_{_{\tmop{MB}}} \left( \tmmathbf{p}_{\bot \tmmathbf{Q}} \right) S
  \left( \tmmathbf{Q}, \tmmathbf{P}_{\| \tmmathbf{Q}} \right), \nonumber
\end{eqnarray}
where in the last line $\mu_{_{\tmop{MB}}} \left( \tmmathbf{p}_{\bot
\tmmathbf{Q}} \right)$ denotes the Maxwell-Boltzmann distribution over
transverse momenta, the Lindblad operators can also be written
\begin{eqnarray}
  L \left( \tmmathbf{p}, \tmmathbf{P}; \tmmathbf{Q} \right) & = &
  \sqrt{\frac{n_{\tmop{gas}}}{m_{\ast}^2}} f \left( \tmop{rel} \left(
  \tmmathbf{p}_{\bot \tmmathbf{Q}}, \mathsf{\tmmathbf{P}}_{\perp \tmmathbf{Q}}
  \right) - \frac{\tmmathbf{Q}}{2}, \tmop{rel} \left( \tmmathbf{p}_{\bot
  \tmmathbf{Q}}, \mathsf{\tmmathbf{P}}_{\perp \tmmathbf{Q}} \right) +
  \frac{\tmmathbf{Q}}{2} \right) \nonumber\\
  &  & \hspace{1em} \times \sqrt{\mu_{_{\tmop{MB}}} \left( \tmmathbf{p}_{\bot
  \tmmathbf{Q}} \right)}^{} \sqrt{S_{} \left( \tmmathbf{Q},
  \mathsf{\tmmathbf{P}} \right)}, \nonumber
\end{eqnarray}
thus putting again into evidence the appearance of the dynamic structure
factor, whose positivity has been exploited in order to take the square root.

The relevant correlation function for the dynamics is thus given by the
Fourier transform with respect to energy of the time-dependent
auto-correlation function of the operator of the bath appearing in Eq.
(\ref{eq:16}). The appearance of the dynamic structure factor has an important
physical meaning, linking the dynamics of the test particle to the density
fluctuations in the medium, as we shall see in Sect.
\ref{sec:fluct-diss-theor}, expressing the molecular, discrete nature of
matter, that is to say one of the basic insights gained by Einstein's
description of Brownian motion.

The master equation (\ref{eq:full}), or (\ref{eq:39}) when considering the
Born approximation, can be seen as a quantum counterpart of the classical
linear Boltzmann equation, as discussed in
{\cite{Petruccione2005a,Hornberger2007c}}, in that it addresses in a quantum
framework the same physical situation described by the classical linear
Boltzmann equation. This is also confirmed by the fact that the diagonal
matrix elements in the momentum representation of the quantum linear Boltzmann
equation do give back the classical linear Boltzmann equation, obviously
expressed with the quantum scattering cross section.

\subsection{Translation-covariant quantum dynamical semigroups}

In Sect. \ref{sec:qlbe} we have considered a test particle interacting through
collisions with a homogeneous background gas. As stressed in Sect.
\ref{sec:transl-invar} the collisions are to be described by an interaction
potential only depending on the relative coordinate, so as not to spoil
invariance under translations. These two requirements lead to a natural
general constraint on the structure of the mapping giving the reduced
dynamics. In fact homogeneity of the bath implies that the statistical
operator describing its equilibrium state commutes with the momentum operator
of the bath. Similarly the considered interaction $V$ ensures that the total
Hamiltonian $H$ commutes with the momentum operator of the whole system, which
we can write as $\tmmathbf{P}_S + \tmmathbf{P}_B$. Let us now consider the
reduced operator of the test particle at time $t$ obtained by taking the trace
over the bath degrees of freedom of the statistical operator of the total
system. Considering a factorized initial state one has
\begin{eqnarray}
  \mathcal{U}_t \left[ \varrho_S \right] & = & \tmop{Tr}_B \left( e^{-
  \frac{i}{\hbar} Ht} \varrho_S \otimes \varrho_B e^{+ \frac{i}{\hbar} Ht}
  \right) . \nonumber
\end{eqnarray}
Exploiting further the aforementioned constraints
\begin{eqnarray}
  \left[ \varrho_B, \tmmathbf{P}_B \right] = 0 \hspace{2em} & \tmop{and} &
  \hspace{2em} \left[ H, \tmmathbf{P}_S + \tmmathbf{P}_B \right] = 0,
  \nonumber
\end{eqnarray}
one immediately has, for any vector $\tmmathbf{a} \in \mathbbm{R}^3$
\begin{eqnarray}
  \tmop{Tr}_B \left( e^{- \frac{i}{\hbar} Ht} e^{- \frac{i}{\hbar}
  \tmmathbf{P}_S \cdot \tmmathbf{a}} \varrho_S e^{+ \frac{i}{\hbar}
  \tmmathbf{P}_S \cdot \tmmathbf{a}} \otimes \varrho_B e^{+ \frac{i}{\hbar}
  Ht} \right) & = & e^{- \frac{i}{\hbar} \tmmathbf{P}_S \cdot \tmmathbf{a}}
  \tmop{Tr}_B \left( e^{- \frac{i}{\hbar} Ht} \varrho_S \otimes \varrho_B e^{+
  \frac{i}{\hbar} Ht} \right) e^{+ \frac{i}{\hbar} \tmmathbf{P}_S \cdot
  \tmmathbf{a}}, \nonumber
\end{eqnarray}
so that a mapping $\mathcal{U}_t$ giving the reduced dynamics must obey
\begin{eqnarray}
  \mathcal{U}_t \left[ e^{- \frac{i}{\hbar} \tmmathbf{P}_S \cdot \tmmathbf{a}}
  \varrho_S e^{+ \frac{i}{\hbar} \tmmathbf{P}_S \cdot \tmmathbf{a}} \right] &
  = & e^{- \frac{i}{\hbar} \tmmathbf{P}_S \cdot \tmmathbf{a}}  \mathcal{U}_t
  \left[ \varrho_S \right] e^{+ \frac{i}{\hbar} \tmmathbf{P}_S \cdot
  \tmmathbf{a}} . \nonumber
\end{eqnarray}
This condition is known as covariance under translations. Focusing on the
Hilbert space of the massive test particle considered in the present paper,
that is to say $L^2 \left( \mathbbm{R}^3 \right)$, the condition can be stated
as follows. Given the unitary representation $\mathsf{U} ( \mathbf{a}) = \exp
(- i \mathbf{a} \cdot \mathsf{P / \hbar})$, $\mathbf{a} \in \mathbbm{R}^3$ of
the group of translations $\mathbbm{R}^3$ in $L^2 \left( \mathbbm{R}^3
\right)$, a mapping $\mathcal{L}$ acting on the statistical operators in this
space is said to be translation-covariant if it commutes with the action of
the unitary representation, i.e.
\begin{equation}
  \label{eq:33} \mathcal{L} [ \mathsf{U} ( \mathbf{a}) \varrho
  \mathsf{U}^{\dagger} ( \mathbf{a})] = \mathsf{U} ( \mathbf{a}) \mathcal{L}
  [\varrho] \mathsf{U}^{\dagger} ( \mathbf{a}),
\end{equation}
for any statistical operator $\varrho$ and any translation $\mathbf{a}$. The
general structure of generators of quantum dynamical semigroups complying with
this covariance condition has been obtained by Holevo
{\cite{Holevo1993a,Holevo1993b}}, and it turns out that the requirement of
translation covariance puts very stringent constraints on the Lindblad
operators appearing in the expression of the generator. These results, while
obviously fitting in the general framework set by the famous Lindblad result
{\cite{Lindblad1976a,Gorini1976a}}, go beyond it giving much more detailed
information on the possible choice of operators appearing in the Lindblad
form, information conveyed by the symmetry requirements and relying on a
quantum generalization of the L\'evy-Khintchine formula. They therefore also
provide a precious starting point for phenomenological approaches exploiting
relevant physical symmetries. Referring to the papers by Holevo for the
related mathematical details (see also {\cite{Vacchini2005b}} for a brief
r\'esum\'e), the generator can be expressed as
\begin{equation}
  \label{eq:34} \mathcal{L} [\varrho] = - \frac{i}{\hbar} \left[ H \left(
  \mathsf{P} \right), \varrho \right] + \mathcal{L}_G [\varrho] +
  \mathcal{L}_P [\varrho],
\end{equation}
with $H ( \mathsf{P})$ a self-adjoint operator which is only a function of the
momentum of the test particle. The so-called Gaussian part $\mathcal{L}_G$ is
given by
\begin{equation}
  \label{eq:35} \mathcal{L}_G [\varrho] = - \frac{i}{\hbar} \left[
  \mathsf{Y}_0 + H_{\mathrm{\tmop{eff}}} ( \mathsf{X}, \mathsf{P}), \varrho
  \right] + \sum_{k = 1}^r \left[ K_k \varrho K_k^{\dagger} - \frac{1}{2}
  \left\{ K_k^{\dagger} K_k, \varrho \right\} \right],
\end{equation}
where
\begin{eqnarray}
  K_k = \mathsf{Y}_k + L_k ( \mathsf{P}), &  & \mathsf{Y}_k = \sum_{i = 1}^3
  a_{ki} \mathsf{X}_i, \hspace{1em} H_{\mathrm{\tmop{eff}}} ( \mathsf{X},
  \mathsf{P}) = \frac{\hbar}{2 i} \sum_{k = 1}^r ( \mathsf{Y}_k L_k (
  \mathsf{P}) - L_k^{\dagger} ( \mathsf{P}) \mathsf{Y}_k) \nonumber
\end{eqnarray}
with $k = 0, \ldots, r \leq 3$ and $a_{ki} \in \mathbbm{R}$, while the
remaining Poisson part takes the form
\begin{equation}
  \label{eq:36} \mathcal{L}_P [\varrho] = \int d \mu (\tmmathbf{Q}) \sum_{j =
  1}^{\infty} \left[ e^{\frac{i}{\hbar} \tmmathbf{Q} \cdot \mathsf{X}} L_j
  (\tmmathbf{Q}, \mathsf{P}) \varrho L^{\dagger}_j (\tmmathbf{Q}, \mathsf{P})
  e^{- \frac{i}{\hbar} \tmmathbf{Q} \cdot \mathsf{X}} - \frac{1}{2} \left\{
  L^{\dagger}_j (\tmmathbf{Q}, \mathsf{P}) L_j (\tmmathbf{Q}, \mathsf{P}),
  \varrho \right\} \right],
\end{equation}
with $d \mu (\tmmathbf{Q})$ a positive measure. The names Gaussian and Poisson
arise in connection with the different contributions in the classical
L\'evy-Khintchine formula {\cite{Feller1971}}. In the Gaussian part the
$\mathsf{Y}_k$ are linear combinations of the three position operators of the
test particle, while the generally complex functions $L_k ( \mathsf{P})$ have
an imaginary part accounting for friction, typically given by a linear
contribution, corresponding to a friction term proportional to velocity. In
the Poisson part a continuous index $\tmmathbf{Q}$ appears, together with the
usual sum over a discrete index $j$. The expression is characterized by the
appearance of the unitary operators $\exp (i\tmmathbf{Q} \cdot \mathsf{X} /
\hbar)$, expressing momentum kicks, and of the functions $L_j (\tmmathbf{Q},
\mathsf{P})$, operator-valued in that they depend on the momentum operators of
the test particle $\mathsf{P}$.

\section{Fluctuation-dissipation theorem}\label{sec:fluct-diss-theor}

As already stressed the two-point correlation function appearing
operator-valued in the master equation is the dynamic structure factor
(\ref{eq:37}), where the Fourier transform of the number-density operator
$\rho_{\tmmathbf{Q}}$, as given in (\ref{eq:15}), appears. This function is
directly related to the density fluctuations in the medium, as it can be seen
writing it in the following way {\cite{Lovesey1984}}:
\begin{equation}
  \label{eq:49} S (\tmmathbf{Q}, E) = \frac{1}{2 \pi \hbar} \int dt \int d^3
  \hspace{-0.25em} \tmmathbf{x} \hspace{0.25em} e^{\frac{i}{\hbar} (Et
  -\tmmathbf{Q} \cdot \tmmathbf{x})} \frac{1}{N} \int d^3 \hspace{-0.25em}
  \tmmathbf{y} \hspace{0.25em} \left\langle N_B (\tmmathbf{y}) N_B
  (\tmmathbf{x}+\tmmathbf{y}, t) \right\rangle,
\end{equation}
i.e., as Fourier transform with respect to energy and momentum transfer of the
time dependent density correlation function . Here the connection with density
fluctuations and therefore discrete nature of matter is manifest. Introducing
the real correlation functions
\begin{equation}
  \label{eq:51} \phi^- (\tmmathbf{Q}, t) = \frac{i}{\hbar N} \langle
  [\rho_{\tmmathbf{Q}} (t), \rho_{\tmmathbf{Q}}^{\dagger}] \rangle
  \hspace{2em} \mathrm{\tmop{and}} \hspace{2em} \phi^+ (\tmmathbf{Q}, t) =
  \frac{1}{\hbar N} \langle \{\rho_{\tmmathbf{Q}} (t),
  \rho_{\tmmathbf{Q}}^{\dagger} \} \rangle,
\end{equation}
where $\{,\}$ denotes the anticommutator, the fluctuation-dissipation theorem
can be formulated in terms of the dynamic structure factor as follows
\begin{eqnarray}
  \phi^- (\tmmathbf{Q}, t) & = & - \frac{2}{\hbar} \int^0_{- \infty} dE
  \hspace{0.25em} \sin \left( Et / \hbar \right) \left( 1 - e^{\beta E}
  \right) S (\tmmathbf{Q}, E)  \label{eq:52}\\
  \phi^+ (\tmmathbf{Q}, t) & = & - \frac{2}{\hbar} \int^0_{- \infty} dE
  \hspace{0.25em} \cos \left( Et / \hbar \right) \coth \left( \beta / 2 E
  \right) \left( 1 - e^{\beta E} \right) S (\tmmathbf{Q}, E) . \nonumber
\end{eqnarray}
We recall that contrary to the usual perspective in linear response theory we
take as positive momentum and energy transferred to the particle. The dynamic
structure factor can also be directly related to the dynamic response function
$\chi'' (\tmmathbf{Q}, E)$ {\cite{Pitaevskii2003}}, according to
\begin{equation}
  \label{eq:53} S (\tmmathbf{Q}, E) = \frac{1}{2 \pi} \left[ 1 - \coth \left(
  \frac{\beta}{2} E \right) \right] \chi'' (\tmmathbf{Q}, E) = \frac{1}{\pi}
  \frac{1}{1 - e^{\beta E}} \chi'' (\tmmathbf{Q}, E),
\end{equation}
the relationship leading to the important fact that while the dynamic response
function is an odd function of energy, the dynamic structure factor obeys the
so-called detailed balance condition
\begin{equation}
  \label{eq:54} S (\tmmathbf{Q}, E) = e^{- \beta E} S (-\tmmathbf{Q}, - E),
\end{equation}
a property granting the existence of a stationary state for the master
equation {\cite{Vacchini2001b}}.

The significance of the appearance of the dynamic structure factor in
connection to the so-called fluctuation-dissipation theorem is to be traced
back to a seminal paper by van Hove {\cite{VanHove1954,Schwabl2003}}. In fact
he showed that the scattering cross-section of a microscopic probe off a
macroscopic sample can be written in Born approximation
\begin{equation}
  \label{eq:56} \frac{d^2 \sigma}{d \Omega_{P'} dE_{P'}} (\tmmathbf{P}) =
  \left( 2 \pi \hbar \right)^6 \left( \frac{M}{2 \pi \hbar^2} \right)^2
  \frac{P'}{P} | \tilde{V} (\tmmathbf{Q}) |^2 S (\tmmathbf{Q}, E),
\end{equation}
where a particle of mass $M$ changes its momentum from $\tmmathbf{P}$ to
$\tmmathbf{P}' =\tmmathbf{P}+\tmmathbf{Q}$ scattering off a medium with
dynamic structure factor $S (\tmmathbf{Q}, E)$. This can be seen as a
formulation of the fluctuation-dissipation relationship for the case of a test
particle interacting through collisions with a macroscopic fluid. The energy
and momentum transfer to the particle, characterized by the expression of the
scattering cross-section at l.h.s. of (\ref{eq:56}) are related to the density
fluctuations of the macroscopic fluid appearing through the dynamic structure
factor at r.h.s. of (\ref{eq:56}). One of the basic ideas of Einstein's
Brownian motion, i.e., the discrete nature of matter, once again appears in
the formulation (\ref{eq:56}) of the fluctuation-dissipation relationship.

\section{Friction coefficient for quantum description of Brownian
motion}\label{sec:quant-descr-einst}

We now come to the master equation for the quantum description of Einstein's
Brownian motion. The requirement of translational invariance has been settled
in Sect. \ref{sec:ti}, while the connection between reduced dynamics of the
test particle and density fluctuations in the medium, coming about because of
its discrete nature, has been taken into account in Sect.
\ref{sec:fluct-diss-theor}. The last step to be taken is to consider the test
particle much more massive than the particles making up the gas, i.e., the
Brownian limit $m / M \ll 1$, which in turn implies considering both small
energy and momentum transfers, similarly to the classical case
{\cite{Uhlenbeck1948a,Vacchini2001b}}. We therefore start from (\ref{eq:39})
and consider a free gas of Maxwell-Boltzmann particles, so that taking the
limiting expression of (\ref{eq:41}) when the ratio between the masses is much
smaller than one leads, of necessity as can be seen from the Gaussian
contribution in Holevo's result (\ref{eq:35}) but also from previous work
{\cite{Lindblad1976b,Sandulescu1987a,Diosi1995a}}, to a Caldeira Leggett type
master equation, however without shortcomings related to the lack of
preservation of positivity of the statistical operator. The master equation
takes the form
\begin{equation}
  \label{eq:67} \frac{d \varrho}{dt} = - \frac{i}{\hbar} \left[ \mathsf{H}_0,
  \varrho \right] - \frac{i}{\hbar} \frac{\eta}{2} \sum_{i = 1}^3 \left[
  \mathsf{X}_i, \left\{ \mathsf{P}_i, \varrho \right\} \right] -
  \frac{D^{}_{pp}}{\hbar^2} \sum_{i = 1}^3 \left[ \mathsf{X}_i, \left[
  \mathsf{X}_i, \varrho \right] \right] - \frac{D_{xx}}{\hbar^2} \sum_{i =
  1}^3 \left[ \mathsf{P}_i, \left[ \mathsf{P}_i, \varrho \right] \right],
\end{equation}
with
\begin{equation}
  \label{eq:68} D_{pp} = \frac{M}{\beta} \eta \hspace{1em} \text{and}
  \hspace{1em} D_{xx} = \frac{\beta \hbar^2}{16 M} \eta .
\end{equation}
The friction coefficient $\eta$ is uniquely determined on the basis of the
microscopic information on interaction potential and correlation function of
the macroscopic system, according to
\begin{equation}
  \label{eq:69} \eta = \frac{\beta}{2 M} \frac{2 \pi}{\hbar} (2 \pi \hbar)^3
  n_{\tmop{gas}} \int d^3 \hspace{-0.25em} \tmmathbf{Q} \hspace{0.25em} |
  \tilde{V} (\tmmathbf{Q}) |^2 \hspace{0.25em} \frac{Q^2}{3} S (\tmmathbf{Q},
  E = 0),
\end{equation}
the factor 3 being related to the space dimensions, thus proving in a specific
physical case of interest the so-called standard wisdom expecting the
decoherence and dissipation rate to be connected with the value at zero energy
of some suitable spectral function {\cite{Alicki2004a}}. The transition from
the general master equation (\ref{eq:39}) to the approximate expression
(\ref{eq:67}) has been considered in detail in {\cite{Vacchini2007d}}, where
the microscopic expression for the friction coefficient has been worked out in
detail for the case of a constant scattering cross section. In order to point
out the connection with classical Brownian motion as described by Einstein we
stress that (\ref{eq:67}) is a quantum version of the classical Kramer's
equation. This can be easily seen considering the usual correspondence rules
between classical and quantum mechanics, sending position to multiplication by
the variable and momentum to derivation, or also considering the expression of
(\ref{eq:67}) for the Wigner function, which reads
\begin{eqnarray}
  \frac{\partial}{\partial t} W \left( \tmmathbf{X}, \tmmathbf{P} \right) & =
  & - \frac{\tmmathbf{P}}{M} \cdot \nabla_{\tmmathbf{X}} W \left(
  \tmmathbf{X}, \tmmathbf{P} \right) \nonumber\\
  &  & + \eta \nabla_{\tmmathbf{P}} \cdot \left( \tmmathbf{P} W \left(
  \tmmathbf{X}, \tmmathbf{P} \right) \right) + D_{pp} \Delta_{\tmmathbf{P}} W
  \left( \tmmathbf{X}, \tmmathbf{P} \right) + D_{xx} \Delta_{\tmmathbf{X}} W
  \left( \tmmathbf{X}, \tmmathbf{P} \right), \nonumber
\end{eqnarray}
and in the strong friction limit leads to the classical Smoluchowski equation
with a small quantum correction {\cite{Vacchini2002a}}.

\subsubsection*{Acknowledgements}

The work was partially supported by the Italian MIUR under PRIN05 (BV).


\begin{thebibliography}{10}
  \bibitem[1]{Breuer2007}H.P. Breuer, F.~Petruccione, \tmtextit{The Theory of
  Open Quantum Systems} (Oxford University Press, Oxford, 2007)
  
  \bibitem[2]{Petruccione2005a}F.~Petruccione, B.~Vacchini, Phys. Rev.~E
  \tmtextbf{71}, 046134 (2005)
  
  \bibitem[3]{Lovesey1984}S.~Lovesey, \tmtextit{Theory of neutron scattering
  from condensed matter. Vol.1. Nuclear scattering} (Clarendon Press, Oxford,
  UK, 1984)
  
  \bibitem[4]{Pitaevskii2003}L.~Pitaevskii, S.~Stringari,
  \tmtextit{Bose-Einstein condensation} (Oxford University Press, Oxford,
  2003)
  
  \bibitem[5]{Vacchini2000a}B.~Vacchini, Phys. Rev. Lett. \tmtextbf{84}, 1374
  (2000)
  
  \bibitem[6]{Vacchini2001a}B.~Vacchini, Phys. Rev.~E \tmtextbf{63}, 066115
  (2001)
  
  \bibitem[7]{Vacchini2001b}B.~Vacchini, J.~Math. Phys. \tmtextbf{42}, 4291
  (2001)
  
  \bibitem[8]{Hornberger2006b}K.~Hornberger, Phys. Rev. Lett. \tmtextbf{97},
  060601 (2006)
  
  \bibitem[9]{Hornberger2007c}K.~Hornberger, B.~Vacchini (2008), to appear in
  Phys. Rev. A [arXiv:quant-ph/0711.3109]
  
  \bibitem[10]{Holevo1993a}A.S. Holevo, Rep. Math. Phys. \tmtextbf{32}(2), 211
  (1993)
  
  \bibitem[11]{Holevo1993b}A.S. Holevo, Rep. Math. Phys. \tmtextbf{33}(1-2),
  95 (1993)
  
  \bibitem[12]{Lindblad1976a}G.~Lindblad, Comm. Math. Phys. \tmtextbf{48}(2),
  119 (1976)
  
  \bibitem[13]{Gorini1976a}V.~Gorini, A.~Kossakowski, E.C.G. Sudarshan,
  J.~Math. Phys. \tmtextbf{17}(5), 821 (1976)
  
  \bibitem[14]{Vacchini2005b}B.~Vacchini, Int. J. Theor. Phys.
  \tmtextbf{44}(7), 1011 (2005)
  
  \bibitem[15]{Feller1971}W.~Feller, \tmtextit{An introduction to probability
  theory and its applications. Vol. II} (John Wiley \& Sons Inc., New York,
  1971)
  
  \bibitem[16]{VanHove1954}L.~Van Hove, Phys. Rev. \tmtextbf{95}, 249 (1954)
  
  \bibitem[17]{Schwabl2003}F.~Schwabl, \tmtextit{Advanced quantum mechanics},
  2nd~edn. (Springer, New York, 2003)
  
  \bibitem[18]{Uhlenbeck1948a}C.S.W. Chang, G.E. Uhlenbeck, \tmtextit{The
  kinetic theory of gases}, in \tmtextit{Studies in statistical mechanics},
  edited by J.D. Boer (North-Holland, Amsterdam, 1970), Vol.~5
  
  \bibitem[19]{Lindblad1976b}G.~Lindblad, Rep. Mat. Phys. \tmtextbf{10}(3),
  393 (1976)
  
  \bibitem[20]{Sandulescu1987a}A.~S ndulescu, H.~Scutaru, Ann. Physics
  \tmtextbf{173}(2), 277 (1987)
  
  \bibitem[21]{Diosi1995a}L.~Di\'osi, Europhys. Lett. \tmtextbf{30}, 63 (1995)
  
  \bibitem[22]{Alicki2004a}R.~Alicki, Open Syst. Inf. Dyn. \tmtextbf{11}(1),
  53 (2004)
  
  \bibitem[23]{Vacchini2007d}B.~Vacchini, K.~Hornberger, Eur. Phys. J.~ST
  \tmtextbf{151}, 59 (2007)
  
  \bibitem[24]{Vacchini2002a}B.~Vacchini, Phys. Rev.~E \tmtextbf{66}(2),
  027107 (2002)
\end{thebibliography}
\end{document}